\documentclass{cernyrep} 
\usepackage{texnames}
\usepackage[T1]{fontenc}
\usepackage{epsf,epsfig,subfigure,graphicx,amsmath,amssymb}
\usepackage{color}
\usepackage{mathrsfs}
\usepackage{verbatim}
\begin{document}
\title{Beyond the Standard Model}
\author{Mihoko M. Nojiri}
\institute{Theory Center, IPNS, KEK, Tsukuba, Japan, and 
Kavli IPMU, The University of Tokyo, Kashiwa, Japan}

\maketitle
\begin{abstract}
A Brief review on the physics beyond the Standard Model.
\end{abstract}
\section{Quest of BSM }

Although  the standard model of elementary particles(SM) describes the 
high energy phenomena very well,  particle physicists  have been attracted 
by the physics beyond the Standard Model (BSM). There are very good 
reasons about this; 

\begin{enumerate}
\item The SM Higgs sector is not natural. 
\item There  is no dark matter candidate in the SM. 
\item  Origin of three gauge interactions is not understood in the SM. 
\item   Cosmological observations suggest  an inflation period in the early universe. 
The non-zero baryon number of our universe is not consistent 
with the inflation picture unless a new interaction is introduced. 
\end{enumerate}

The Higgs boson  candidate was discovered recently.  The study of the Higgs boson nature  is extremely important 
for the BSM study. 

The Higgs boson is a spin 0 particle, and  
the  structure of the radiative correction is quite different from those of 
fermions and gauge bosons.   The correction of the Higgs boson mass 
is proportional to the cut-off scale, called ``quadratic divergence". If the cut-off scale is high, 
the correction becomes unacceptably large compared with the on-shell mass of the Higgs boson. 
This is often called a ``fine turning problem".
Note that such quadratic divergence does not appear in the radiative 
correction to the fermion and gauge boson masses. They  are protected 
by the chiral and gauge symmetries, respectively. 

The problem can be solved  if there are an intermediate scale  where new particles appears,  and  the radiative correction from the new particles compensates the SM radiative correction. 
The scale is probably much less than  $\mathcal{O}$(100)~TeV,  where the ratio  between the SM radiative correction  and the Higgs vev is more  than  1000.  The turning of the factor 1000 may sound unnatural,  but it is  much better than the scale among other parameters, such as Planck scale  to the order of electroweak symmetry breaking, or the large difference among Yukawa couplings.

An idea  to introduce a new particle  that couples to the Higgs boson to cancel one loop level quadratic correction, is not successful, because such accidental cancellation does not  hold all order in the perturbation theory. 
One needs new symmetry to cancel the quadratic divergence in the SM  by a new physics contribution. 
The known ideas to achieve  the reduction of quadratic divergence are
the following; 
\begin{enumerate}
\item  {\bf Supersymmetry}: Extend the SM so that the 
theory has ``supersymmetry". Supersymmetry is the symmetry between 
bosons and fermions, which allows the divergence of  Higgs boson mass controlled by 
 "chiral symmetry" of fermions.  Due to the cancellation among various diagrams involving 
SM particles and their superpartners (SUSY particles), there are no quadratic divergence to the Higgs bosons mass in this theory.  
\item {\bf Dynamical symmetry breaking}: In this theory, a new strong interaction causes the 
spontaneous  gauge symmetry breaking of the SM. The Higgs doublet is a  Nambu-Goldstone boson  of the symmetry breaking and   bound states of fermions charged under the strong interaction,  corresponding  to the  pions in the QCD. The Higgs boson  does not exist above the symmetry breaking scale, so there are no problem of quadratic divergence. 
\item {\bf Extra dimension} Although we recognize that we live in the four dimensional space-time, 
we might live in more than the five dimension space time where the extra dimensions are compactified. 
The true Planck scale may be much closer to electroweak scale in such a theory, or the 
fundamental parameters in the Higgs sector  is of the order of Planck scale in the higher dimensional theory but looks small in the effective four dimensional theory.  In some class 
of the model the Higgs boson may be a part of gauge boson in the 
5th dimension so that the divergence of the Higgs mass parameters is controlled by the gauge symmetry. 
\end{enumerate}

Those models are constrained strongly by precision measurements.  Currently 
there are no measurements with significant deviation from the SM predictions. 
In the SM theory, one can predict various observables  from a few fundamental parameters: the  gauge couplings $g_i( i=1,2,3) $, and  the Higgs vacuum expectation value (vev)  $v$. 
By measuring the deviations from the SM predictions, we can set constrains on  the new physics.  
Especially, the $S$ and $T$ parameters which parametrize the new physics 
contributions to the gauge two point functions 
are sensitive to all particles that couple to the gauge bosons.  Measurements of flavor changing neutral current (FCNC) constrain the existence of flavor off-diagonal interactions. 
Very precisely measured parameters sometimes exhibit  significant  deviations from the SM predictions. Currently muon  anomalous magnetic moment deviates from the SM prediction 
by more than 3 $\sigma$.  It  is  sensitive to the new physics that couples to muon. 

The quadratic divergence of the Higgs sector exists if the divergence is 
estimated by the momentum cut off  $\Lambda$, the upper bound of the various 
loop integral appearing in the radiative correction in the mass.   
We have to keep it in mind that the quadratic divergence  does not depend on the 
external momentum, therefore it is a regularization  dependent object. 
Especially in dimensional regularization,   quadratic divergence is trivially zero. 
Then, is there any reason that we should take the fine turning problem seriously? 

The fine turning argument based on momentum cut-off is
justified in the case that the theory has large symmetry at some higher energy scale. 
 For example, in the supersymmetric model, the regularization must respect to supersymmetry
 and one cannot subtract all quadratic divergence. To this end, the Higgs sector receives radiative corrections  proportional to the SUSY scale  (superpartner mass scale)  under correct regularization.  
In the limit that superpartners are much heavier than SM particles, the low energy theory looks like the SM with the momentum cutoff at the SUSY scale.  Fine turning arguments 
hold for the theories with an intermediated scale above which a new symmetry emerges. 

There is another indication of the existence of new physics between the weak scale and the Planck scale. 
We may consider the Higgs potential at large field value in the SM and study the 
stability. The potential is a function of the top and Higgs masses, and current top and 
Higgs mass measurements favor metastable Higgs potential.  There is not 
any reason 
that the Higgs vev should fall in such a metastable point, and this also 
suggests that additional  particles that couple to the Higgs sector 
change the shape of the potential. 

Another strong indication of new physics is the existence of dark matter in our Universe. 
Global fit of the cosmological observation favors the existence of stable, neutral particle, dark matter, which accounts for  27\% of the total 
energy of our Universe.  The existence of the dark  matter is also confirmed 
by various  observations of the stellar objects.   Rotation curve of the stars of the 
galaxy indicates that galaxies are dominated by the non-luminous component. 
The is also a technique  to measure the matters extended beyond the 
galaxy scale using gravitational lensing.  

Our universe is $1.38\times 10^{10}$ years old, roughly $10^{17}$~s $\leftrightarrow$  $10^{-43}$ GeV$^{-1}$. The dark matter life time must be at least of the oder of the age of the Universe 
to remain in the current Universe. \footnote{
In order to avoid the constraints coming from cosmic ray observations, the lifetime of the dark matter 
in our Universe must be  significantly longer than the age of the Universe. }  On the other hand,  a particle with mass $m$~(GeV) with interaction suppressed by $1/M_{pl}$  
has a decay width of order of $g^2 (m/{\rm 1~GeV})^3 10^{-38}$ GeV.
Namely the lifetime, $\tau\sim g^{-2} 10^{14} {\rm s}/ ({\rm m/1~GeV} )^3 $,
would  be much shorter than the life  of our Universe ($\sim 4.3\times 10^{17}$ s), 
where $g$ is the coupling of the decay vertex.  To account for the lifetime of  the dark matter in our universe,  its decay must be very strongly suppressed, or forbidden.  

For the case of the SM particles, existence of stable particles is ensured by the symmetry. 
Electron is the lightest charged particle and electronic charge is conserved by the 
gauge symmetry. Proton is the lightest bound state of quarks.  There are no interaction to break proton in the SM, because  number of quark is conserved for interaction with the gauge bosons or the Higgs boson, and direct interaction with electron is forbidden by the gauge symmetry.  
It is possible to conserve the Baryon number 1/3 to the  quarks in the SM, and this reflects the
fact that proton is stable.  To consider the particle model involving  the stable (or long-lived) dark matter, we must introduce new symmetry to protect the dark matter from decaying.

Another puzzle of the SM is the  hyper-charge assignments of the fermions. In the first glance, it is not easy to find the rules to assign the charge to the SM matters. But,  it fits  very nicely to the representation of a $SU(5)$ group, where $SU(3)\times SU(2) \times U(1) $ generators are embedded as 
\begin{equation}
T^a_{SU(3)}=\left(\begin{array}{cc}
\lambda^a & 0 \cr 
    0         & 0 
\end{array}
\right)
\  \  \ 
T^i_{SU(2)}=\left( \begin{array}{cc}
 0  & 0 \cr
    0    &      \sigma^i 
\end{array}
\right)  
\ \ \ 
T_{U(1)}=\left( 
\begin{array}{cc} 
-\frac{1}{3}{\bf 1}_3 & 0 \cr 
 0    & \frac{1}{2}{\bf 1}_2
\end{array}
\right). 
\end{equation}
Here, $\lambda^a$ and $\sigma^i$ are the $SU(3)$ and $SU(2)$ generators, {\bf 1}$_3$ and {\bf 1}$_2$ are
$3\times 3$ or $2\times 2$ unit matrix, and $T_{SU(3)}$, $T_{SU(2)}$, $T_{U(1)}$ satisfy 
the commutation relations of $SU(3)$, $SU(2)$, and $U(1)$ generators.  
Under this generator assignment, $\bf{5^*}$ and {\bf 10} representations of $SU(5)$  
have a charge assignment as 
\begin{equation}
{\bf 5^*}=\left(\begin{array}{c}
 (3^*, 1) _{1/3}\cr
(1,2)_{-1/2}  
\end{array}
 \right),
\end{equation}
 while {\bf 10 } representation is decomposed into  
 $(3,2)_{1/6}$ $\oplus$ $(3^*, 1)_{-2/3}$ $\oplus$ $ (1, 1)_1$ 
 which reside in the $5\times 5$ antisymmetric matrix as 
 \begin{equation}
   {\bf 10}=\left(
  \begin{array}{cc}
  (3^*,1)_{-2/3} & (3,2)_{1/6}\cr
  * & (1, 1)_1
 \end{array}
 \right).
 \end{equation}
This suggests that  $SU(3)$ $\times$ $SU(2)$ $\times$ $U(1)$ symmetry of the 
SM can be unified into the $SU(5)$ gauge symmetry. To realize this, 
 the  SM three gauge 
couplings  must unify at the short distance, so that the $SU(5)$ symmetry is recovered
above that scale.  
The gauge couplings at the short distance is calculated by utilizing the SM renormalization group equations  from the low energy inputs.  They  do not unify for the particle content 
of the SM, therefore to realize the idea of GUT, new set of particles are needed. 
We will see a successful gauge coupling unification is realized in the Supersymmetric model
in the next section. 

\section{Supersymmetry}
 Supersymmety is the symmetry exchanging bosons into fermion, and fermions into bosons. 
 The generators of the supersymmetric transformation satisfy  the following anti-commutation relations 
 \begin{equation}\label{susy}
 \left\{Q^{\alpha}, \bar{Q}_{\dot{\beta}} \right\}= 2 \sigma^{\mu}_{\alpha,\dot{\beta}} P_\mu 
 \end{equation}
 Here $Q$ is a spin $1/2$ and mass dimension $1/2$  operator and    $\alpha$ and $\dot{\beta}$ 
 $(=1,2)$  are the  spin indices of chiral and anti-chiral fermions, and 
 $\sigma^{\mu}=(1,\sigma^i)$ is the Pauli matrices.

This anti-commutation relation can be reduced for any  massive eigenstate $\vert a \rangle$ by taking the 
rest frame $P^{\mu}\vert a \rangle= m_a \delta_{0\mu}  \vert a\rangle$ as follows:
\begin{equation}
 \left\{Q^{\alpha}, \bar{Q}_{\dot{\beta}}\right\}= 2 \delta_{\alpha,\dot{\beta}} m_a .
\end{equation} 
 The relation is same as that of a two-fermion system in quantum mechanics. 
 One can construct an irreducible representation of this algebra starting from a 
 state which annihilates any $\bar{Q}_i$. Suppose the state is spin 0, $\vert 0 \rangle$, 
 all possible states  are  generated as follows; 
 \begin{equation}
 \vert 0 \rangle  \rightarrow  Q_1\vert 0 \rangle , Q_2\vert 0 \rangle  \rightarrow
 Q_1 Q_2\vert 0 \rangle.  
 \end{equation}
 Because $Q_1 Q_1 =$ $Q_2 Q_2=0$,  no more state can be obtained 
 by multiplying the generator  $Q_i$.   Two spin 0 states and two spin $1/2$ states 
 are obtained. These states form a SUSY multiplet, and  the spin $0$ states are the superpartners of the spin $1/2$ states and vise versa.  Because this multiplet contains spin 1/2 states, we can 
 regard this as a matter multiplet. 
 
  Starting from a spin 1/2 state annihilating  $\bar{Q}$  
 one gets two  spin $1/2$ fermion states, a spin $1$ massive  bosonic states  and  a spin $0$ bosonic state,  namely 4 fermion degrees of freedom and 4 bosonic degrees of freedom. 
 This may be regarded as two chiral fermions, one massive gauge boson and 
 one massive Higgs boson.  
Repeating similar analysis to the  massless particles,  one obtains states with helicity $h=\lambda$ and $\lambda+1/2$. If $\lambda=1/2$, a massless gauge boson and its superpartner fermion make a supersymmetric multiplet.  The number of bosonic degrees of freedom is the 
 same as that of fermionic degrees of freedom in this theory. 

All states in the above multiplet have the same mass, which 
looks  irrelevant 
for describing real particles, but it is known that such mass degeneracy is removed by 
spontaneous supersymmetry breaking. Supersymmetry breaking is discussed in the next section. 
 
The minimal supersymmetric standard model (MSSM)  is an extension of the SM
that has a supersymmetry in the limit where all particle masses are 
ignored.  The model is thought to be an effective theory of a fully supersymmetic theory. 
Due to  the spontaneous supersymmetry breaking of the  full theory,  the superparters of the SM
 particles receive a mass much higher than the SM particles.  A superpartner of a fermion is called sfermion  and it is a spin 0 particle.  
 A superpartner of a gauge boson is called gaugino and has spin $1/2$.  A Higgs boson superpartner is  called a higgsino and  has spin $1/2$.  
The particle content of the MSSM is given in 
Table~\ref{Table:MSSM}. The SM particles and their superpartners 
have same charge,  because the generator of supersymmetric transformation $Q$ 
commutes with the SM $SU(3)\times SU(2) \times U(1)$ transformation. 
The number of Higgs doublets  is two in the MSSM because  one should  add 
two Higgsinos, chiral fermions with charge $(1, 2)_{\pm 1/2}$ in the SM because of 
a condition of anomaly cancellation.  

\begin{table}
\label{Table:MSSM}
\begin{center}
\caption{Particle content of the Minimal Supersymmetric Standard Model. }
\begin{tabular}{|c |c|c| }
\hline 
represenations & quark  & squark\cr
\hline
  $(3, 2)_{1/6}$ & $q_L=(u,d)_L$ & $\tilde{q}_L= (\tilde{u}_L, \tilde{d}_L)$
\cr 
$(3^*,1)_{-2/3}$ & $u_R^c$ & $(\tilde{u}_R)^c$
\cr
$(3^*,1)_{1/3}$ & $(d_R)^c$ & $(\tilde{d}_R)^c$
 \cr
 \hline 
 & lepton  & slepton \cr
 \hline
  $(1, 2)_{1/2}$ & $l_L=(\nu,e)_L$ & $\tilde{q}_L= (\tilde{\nu}_L, \tilde{e}_L)$
\cr 
$(1,1)_{1}$ & $(e_R)^c$ & $(\tilde{e}_R)^c$
\cr
\hline
 & Higgsino & Higgs \cr 
 \hline
   $(1, 2)_{-1/2}$ & $(\tilde{H}^0_1,\tilde{H}^-_1)$ & $(H^0_1,H^-_1)$ \cr
      $(1, 2)_{1/2}$ & $(\tilde{H}^+_2,\tilde{H}^0_2)$ & $(H^+_2,H^0_2)$ \cr
\hline 
 & spin 1/2   & spin 1 \cr 
         $(8,1)_0$                & $\tilde{G}$ (gluino) & $G^{\mu}$ \cr 
          $(1,3)_0$               & $\tilde{W}$ (wino) & $W^{\mu}$  \cr 
                    $(1,1)_0$               & $\tilde{B}$ (bino) & $B^{\mu}$  \cr 
                    \hline 
\end{tabular}
\end{center}

\end{table}

As one can see from Table~\ref{Table:MSSM}, 
the number of particles are doubled in the MSSM. The supersymmetry specifies all dimensionless couplings of interactions of new particles, such as four point interaction of scalers and Yukawa couplings, while mass parameters of superpartners are undetermined. To understand 
the coupling relations, one needs to understand the supersymmetric field theory. 
In this lecture, I do not have enough time to talk about it in detail, so I just sketch the 
important elements. 

Fields in the same supersymmetric matter multiplet 
can be arranged in a ``chiral superfield'' which is a function of coordinate $x$, 
$\theta$ and $\bar{\theta}$ a grassmanian  Lorentz spinors  with mass dimension $-1/2$,
\begin{equation}
\Phi(x, \theta, \bar{\theta}) =\phi(y) + \sqrt{2}\psi(y) \theta + F(y) \theta \theta ,
\end{equation}
where $y^{\mu}=x^{\mu}-i\theta \sigma^{\mu} \theta$.  Note that 
by redefining the coordinate  from $x$ to $y$,  $\Phi$ becomes a function of 
$y$ and $\theta$, and $\bar{\theta}$ does not appear.  There are only three fields $\phi$, $F$ and $\psi$ appearing as 
the component fields of $\Phi$. When $\theta$ is zero, $\Phi(x) =\phi(x)$, therefore $\Phi$ is an
extension of the scalar field of non-supersymmetric theory. On the other hand, $\Phi(y, \theta) $ represents  both fermonic and bosonic fields simultaneously. 

$\Phi$ is dimension 1, so that  ${\rm dim}(\phi)=1$ and ${\rm dim}(\psi)=3/2$. $F$ is then  spin 0 and dim 2 field. 
The only $dim<4$ kinetic term of $F$ is  $FF^{*}$, therefore $F$ is  not dynamical. 
The product of a chiral superfield is also a chiral superfield depending only $y$ and $\theta$.  On the other hand, 
$\Phi\bar{\Phi}'$ is not a chiral superfield  as it has the terms proportional to $\bar{\theta}$. 

Just as operator  $P^{\mu}$,translation in coordinate space $x$ is expressed as 
$\partial/\partial x$,   
supersymmetric transformation $Q$  is a translation in the $\theta$ and $\bar{\theta}$ space. 
Namely, in the coordinate  representation it is expressed as 
\begin{equation}
S_{\alpha} =\frac{\partial}{\partial \theta^{\alpha}} +
 i(\sigma^{\mu}\partial_{\mu}   \bar{\theta}). 
\end{equation}
The second term
is needed to satisfy the SUSY algebra give in Eq. \ref{susy}.
With this transformation, 
each field transform as 
\begin{eqnarray}
\delta_{SUSY} \phi &=& \sqrt{2} \alpha \psi,\cr 
\delta_{SUSY} \psi &=&  -i\sqrt \partial_{\mu} \sigma^{\mu}  \phi \bar{\alpha} + \sqrt{2} F \alpha, \cr
\delta_{SUSY} F &= &-i \sqrt{2}\bar{\alpha} \partial_{\mu} \bar{\sigma} ^{\mu} \psi, 
 \end{eqnarray}
 where $\alpha$ and $\bar{\alpha}$ are transformation parameters. 
 Under this transformation kinetic term 
 \begin{equation}
{\cal L}_{kin}= \partial_{\mu} \phi \partial^{\mu} \phi^* + i\bar{\psi} \sigma^{\mu} \partial_{\mu} \psi 
+ F^* F 
  \end{equation}
 is invariant. 
 
 There are a few things worth paying attention. First The $\delta_{SUSY} F$ is total derivative. 
 Because the product of chiral superfields is also a superfield, the $\theta\theta$ component 
 $\mathcal{F}$ transforms as $\mathcal{F}= \partial_{\mu} J^{\mu}$, namely $\mathcal{F}$ 
 can be interaction  terms which are invariant under  supersymmetric transformation. 
   For example,  $\Phi_1\Phi_2 \Phi_3$
 gives $F$ term 
 \begin{equation}
{L}_{Yukawa}= F_1\phi_2\phi_3+F_2\phi_1\phi_3 +F_3 \phi_1\phi_2-\psi_1\psi_2\phi_3 
 -\psi_2\psi_3\phi_1-\psi_3\psi_1\phi_2. 
 \end{equation}
 The interaction contains  Yukawa interaction term  $ y_{ijk}\psi_i \psi_j \phi_k$ which is 
 symmetric under the exchange of $i, j, k$, and also the scalar potential  terms proportional to 
 $y_{ijk}F_i \phi_j \phi_k$    Combined with kinetic term $FF^*$, 
interactions of four  point scalar fields  proportional  to $y^2$ is generated. 
The similar relations
also holds for supersymmetric  gauge interactions. The interaction between gaugino-fermion- sfermion is  proportional  the gauge coupling $g$, and there are scalar four point interactions proportional to  $g^2$.  While  many scalar and fermion partners are  introduced, 
there are no new  dimensionless coupling introduced. 

In addition to the $F$ term,  
 $\theta\theta\bar{\theta}\bar{\theta}$ term of general field product, $\mathcal{D}$ 
 is supersymmetric. For example, supersymmetric kinetic term is $\theta\theta\bar{\theta}
 \bar\theta$ term of $\Phi\bar{\Phi}$. 

We now address some important features of supersymmetric models. 

\begin{figure}
\centering\includegraphics[width=5cm, angle=-90]{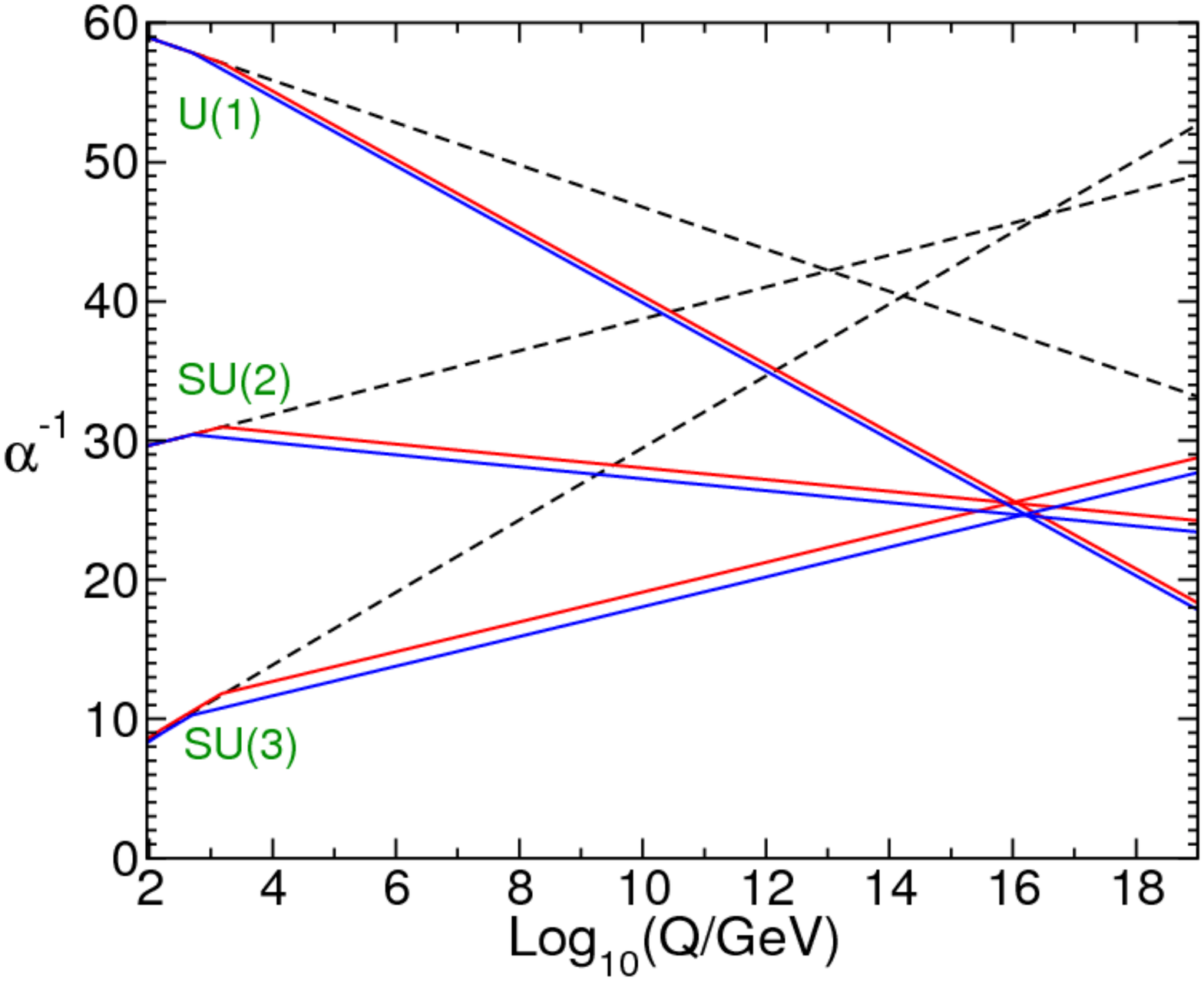}
 \caption{Two loop renormalization group evolution of the gauge couplings in the SM (dashed lines) and MSSM (solid lines)  
 from ``A Supersymmetry primer'' hep-ph/9709356. }\label{fig:unify}
 \end{figure}

\begin{itemize}
\item There are no quadratic divergence in the theory.  The quadratic divergence coming from 
the top loop  is canceled by the stop loop generated by the Higgs-Higgs-stop-stop  four point interaction.  Both of them are proportional to  $y^2$. 
The Higgs four point coupling is proportional to the square of the gauge coupling, and quadratic divergence arising from the diagram  is canceled by the gauge and gaugino-higgsino loops. This is because 
scalar particles are now in a same multiplet with the fermion, and the mass of the fermion
is only logarithmically divergent. The fine-turning in the Higgs sector is now significantly 
reduced. 

\item Because the Higgs four point coupling is a gauge coupling, the Planck scale Higgs 
four point coupling is always positive, therefore significantly  less  in danger of running into 
metastable vacuum  At low energy the Higgs mass is upper bounded by the $Z$ boson 
mass in tree level, and radiative corrections proportional to the $(mt^4/m^2_W) \log (m_{\tilde{t}}/m_t)$ appear in the Higgs boson mass formulae.  This correction is interpreted as 
the running of  the Higgs boson four point coupling from the stop mass scale to the 
top mass scale under the SM renormalization group equation, because below the 
stop mass scale, the theory is effectively the SM. In addition there are 
contribution proportional to the fourth power of stop left-right mixing $X_t$.  See Fig.~\ref{higgscor2} (left)  for the RGE interpretation of the radiative corrections to the Higgs mass. 
In this theory, the Higgs boson mass is calculated from  the 
scalar top mass and  its mixing, therefore the SUSY scale is predicted from the 
Higgs boson mass.  In other words, the measured Higgs boson mass gives a strong constraint to the SUSY mass 
scale and mixing. See Fig.~\ref{higgscor2} (right).
\begin{figure}
\includegraphics[width=7cm, angle=0]{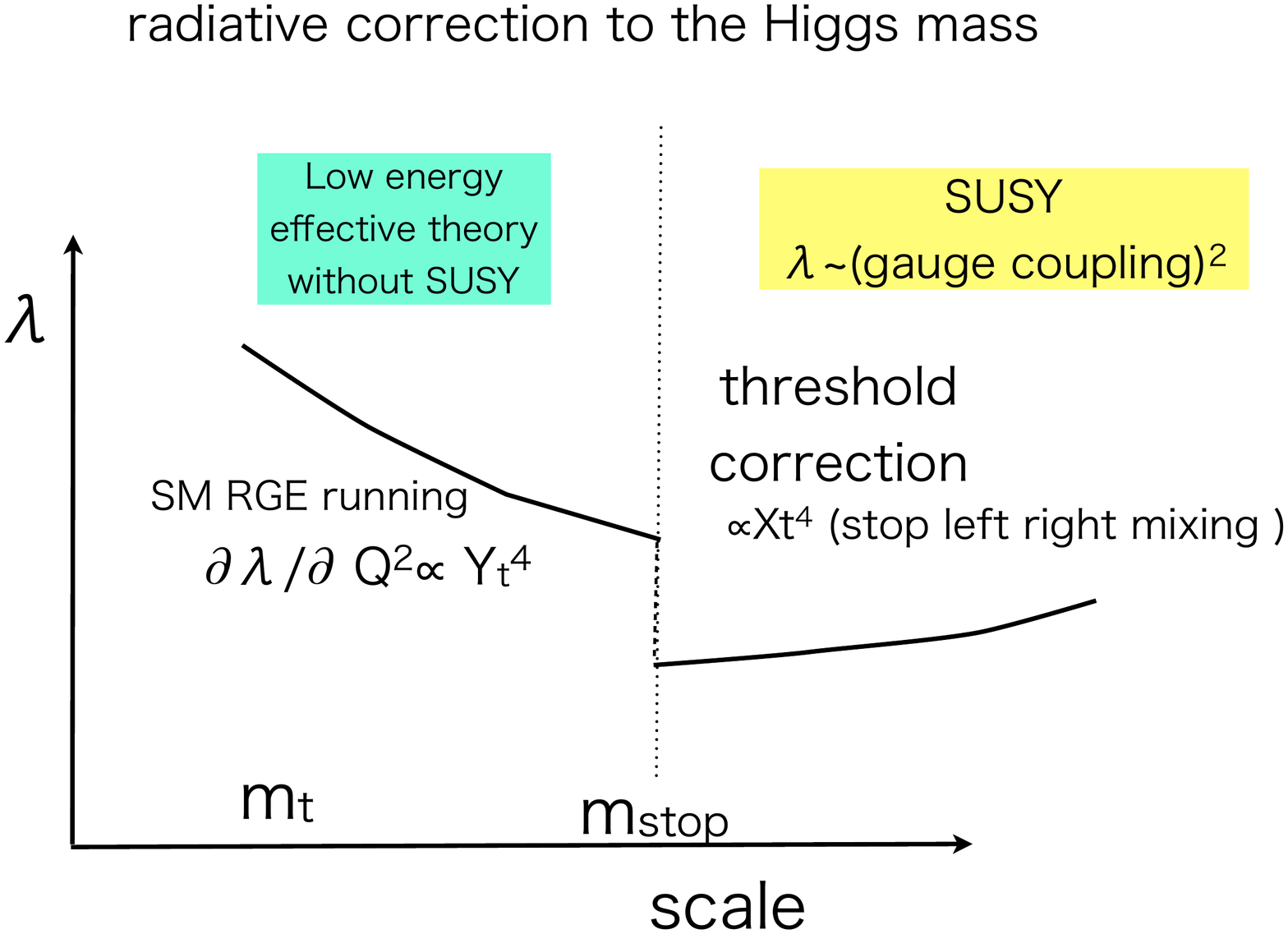}
\includegraphics[width=6cm,angle=90]{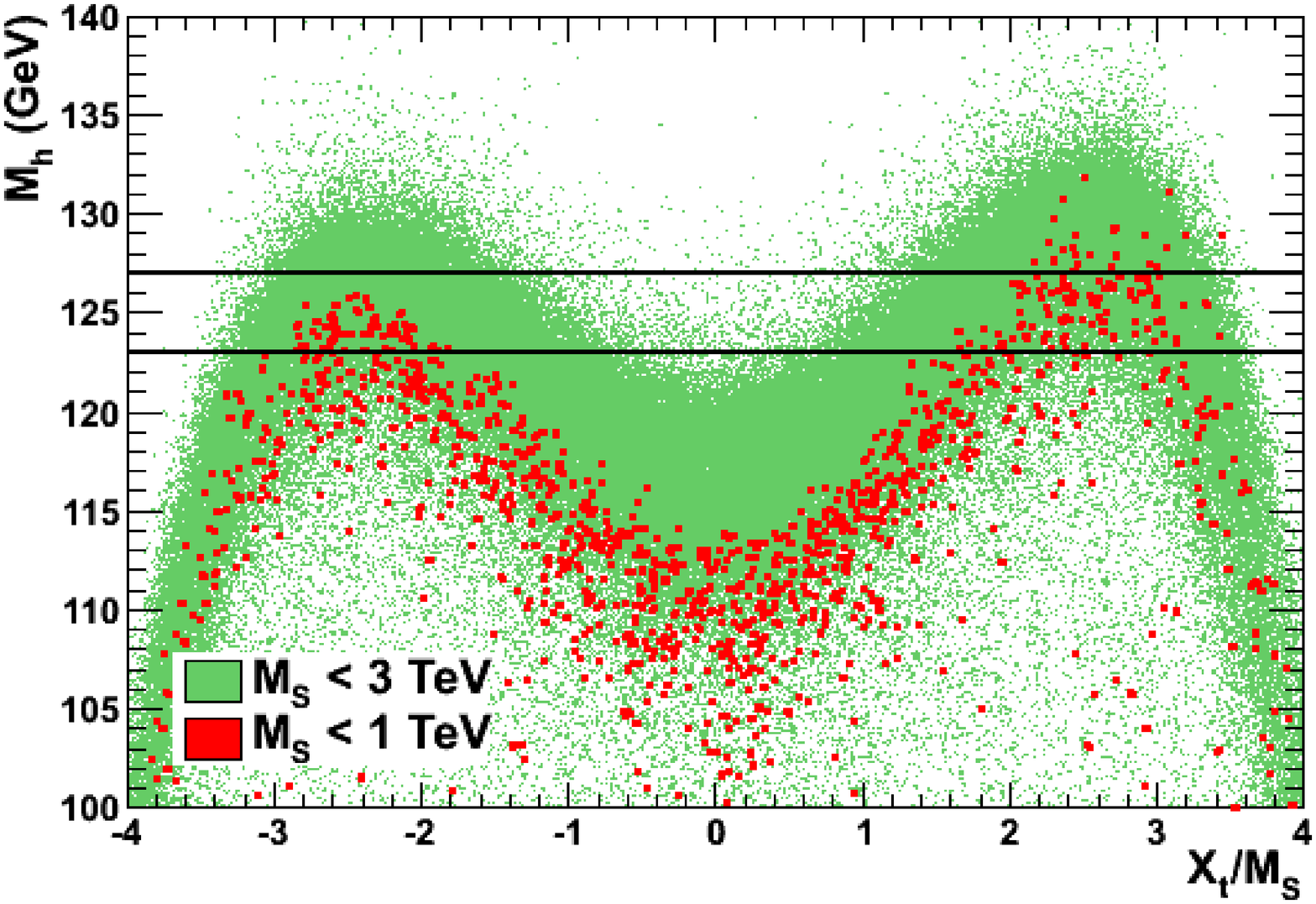}
\caption{Left: the running of Higgs four point coupling changes at the scale of $m_{stop}$. 
Right: Maximal value of the Higgs boson mass as a function of $X_t/M_{SUSY}$ when all the other parameter are scanned. From arXive 1311.0720. } \label{higgscor2}
\end{figure}

\item   In the SM, one cannot write an interaction violating baryon and lepton numbers due to 
the gauge invariance. This is no longer true because Higgsino and lepton doublets have 
same quantum numbers.  The  product of superfields  $W$ whose $\theta \theta$ terms  
  is  the SM Yukawa interactions
 \begin{equation}
W= -y_e H_1 \cdot E^c L- y_d H_1 \cdot D^c Q - y_u H_2 \cdot U^cQ-\mu H_1 \cdot H_2 ,
 \end{equation}
where $Q=\tilde{q}_L+ \theta q_L ...$, $U^c =\tilde{u}^c_R+ \theta u^c_R... $...  are the  superfields whose bosonic component is a sfermion and a fermionic  component is quarks or leptons. 
However,   $\theta\theta$ term of $W'$ 
 \begin{equation}
W' = \epsilon_L  LLE^c+ \epsilon_{BL} L QD^c + \epsilon_B U^cD^cD^c+ \epsilon_{LH} LH_2 
 \end{equation}
  is not forbidden by  the gauge symmetry, because $H_1$ and $L$
   have a same quantum numbers, and 
 $UDD= \epsilon_{abc} U^aD^bD^c$ is a gauge singlet. 
 The interactions violate lepton and/or  baryon numbers  and should be 
 forbidden. 
 
 The symmetry that forbids  $L$ and $B$ violating terms is called  the conserved R-parity. 
 In the MSSM R-parity may be assigned to the superfield and coordinate $\theta $ as follows, 
 \begin{equation}
R(L)=R(E)=R(Q)=R(U)=R(D)=-1, R(H)=1, R(\theta) =-1.   
 \end{equation}
 In this assignment, all the SM particles have $R=1$ and all superpartners  have $R=-1$, and 
 $R(W\vert_{\theta\theta})=1$, and $R(W'\vert_{\theta\theta})=-1$. The 
 interaction term from $W$ multiplicatively conserves $R$ parity, namely, 
 product of $R$ parity of all particles involved in a vertex is one. Namely, $R=-1$ particle 
 decays into the final states  which contains odd number of $R=-1$ particles. If two $R=1$ particle collides, the final state contains even number of $R=-1$ particles. 
By requiring multiplicatively conserved $R$ parity,   the lightest supersymmetric particle (LSP) becomes stable. The LSP  can be a dark matter candidate.    
 
\item  Gauge coupling: In the supersymmetric model, the number of particles is doubled and running  of the gauge couplings would be modified above the SUSY particle mass scale. The gauge couplings unify  at the GUT scale much better than that of the SM as can be seen in Fig.~\ref{fig:unify}. This means "supersymmetric GUT"  is consistent with experimental data, though there are still some fine turning issues when we consider the Higgs sector violating GUT symmetry.  
 \end{itemize}
 
\section{Origin of SUSY breaking}
As we have mentioned already, the MSSM is not 
a complete theory, because it requires  a mechanism to break the supersymmetry 
somewhere outside the MSSM.  A general set up of the SUSY breaking models are  the
following; there are hidden sector $H$, and fields $Z_i$ in the sector $H$ break the supersymmetry  spontaneously. This hidden sector couples to our sector indirectly though a messenger sector.  The particles in the messenger sector have a mass scale $M$.

The spontaneous symmetry breaking is  realized for 
the vacuums which do not annihilate with the supersymmetric generator $Q$ and $\bar{Q}$. 
If such a vacuum exists, there are some fermions $\psi$ whose supersymmetric transformation 
$\delta_{SUSY} \psi = \left\{Q,  \psi\right\}$  has non-zero vev, namely   
$\langle 0 \vert \delta_{SUSY}\phi \vert 0 \rangle =$
$ -\sqrt{2} \langle 0 \vert F \vert 0 \rangle \neq 0$. Some of the superfields in the Hidden section must have non-zero $F$ terms in our setup.  

If $F$ term of $Z$ has  non zero vev,  $\langle Z \rangle=\langle F_Z\rangle \theta\theta$, various mass terms are induced in the low energy effectively.  
A simple example is $\theta\theta\bar{\theta}\bar{\theta} $ term of 
$Z \bar{Z}\Phi\bar{\Phi}/M^2$, which may be  induced through the 
messenger interactions. After the symmetry breaking  the term $(\langle F\rangle ^2/M^2) 
\phi\phi^*$  is the effective SUSY breaking mass term of the scaler boson $\phi$.

There are already severe constraints to the interaction of the messenger sector 
to the MSSM  sector. These constrains come from the flavor changing neutral currents 
such as $K^0$-$\bar{K}^0$ mixing. 
The constraints typically require 
\begin{equation}
\left[
\frac{10{\rm TeV} }{m_{\tilde{q},\tilde{g}} }
\right]^2
\left[ 
\frac{\Delta m^2_{\tilde{q}_{12}}/m^2}{0.1}
\right]^2
<1,
\end{equation}
where $m^2_{\tilde{q}_{12}}$ is a mixing parameter of the first and second generation 
squark,  and $m^2$ is diagonal squark masses. 
The SUSY breaking sector $H$ therefore must couple to the MSSM matter sector universally.
\begin{figure}
\centering\includegraphics[width=7cm, angle=0]{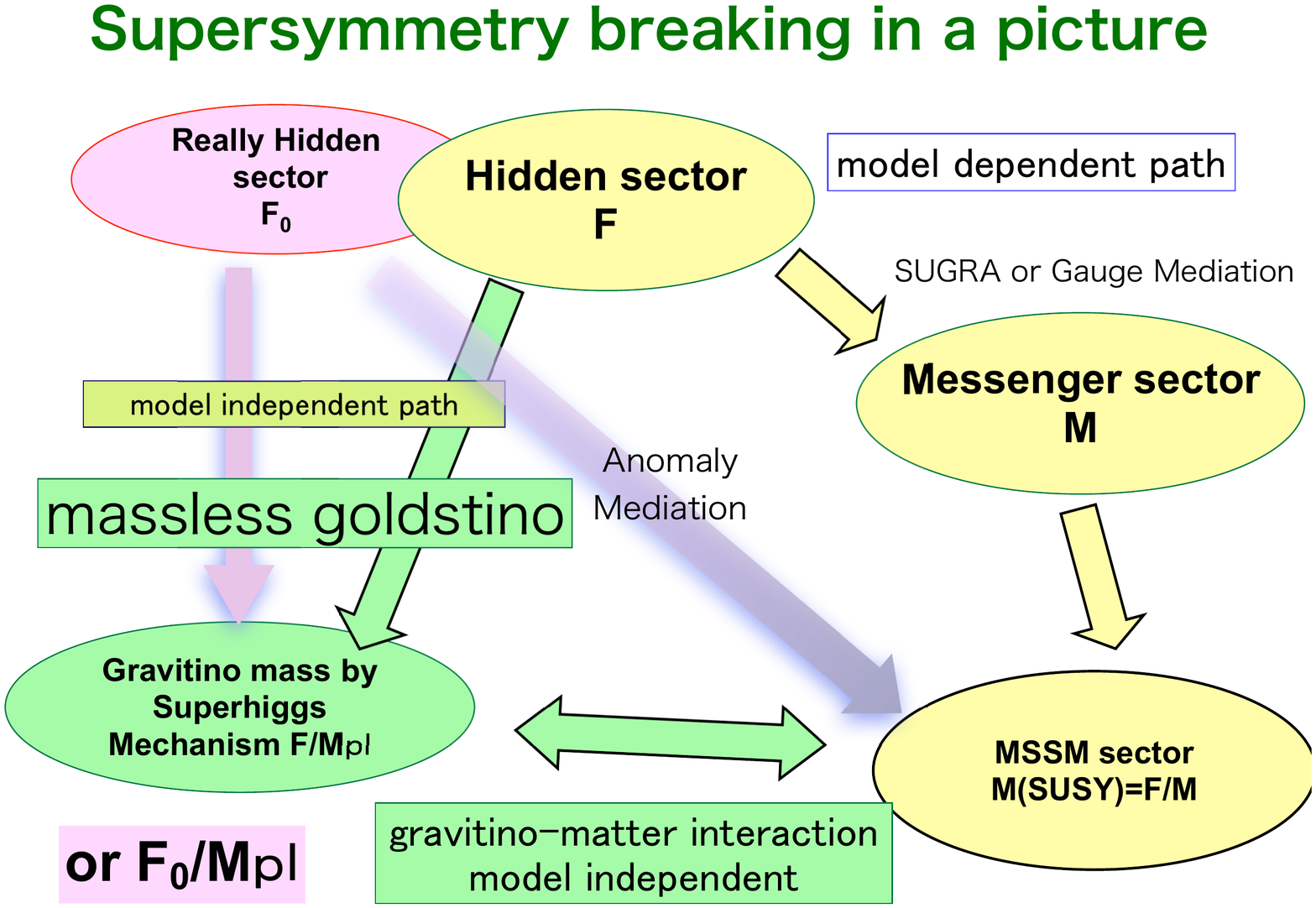}
\caption{Relation between the  MSSM sector and SUSY breaking sector.}\label{SUSYB}
\end{figure}

Several mechanisms have been proposed to assure the universality of the soft scalar masses. 
The supergravity model uses the gravity interaction as the messenger mechanism, 
on the other hand, gauge mediation models uses some vector-like matters charged under the SM
gauge groups as the messenger fields.  Even if there are no direct couplings between the 
MSSM and SUSY breaking sectors, there are mediation mechanism through the 
superconformal anomaly, and the model utilizing this is called anomaly mediation model.  

It is difficult to access the HIdden sector directly.  
The   SUSY breaking of the total theory $F_0$  and 
mass of the gravitino(super partner of graviton)  $m_{3/2}$ is related as $m_{3/2}=F_0/M_{pl}$. 
The gravitino could be the LSP, in that case the next lightest SUSY particle(NLSP) is long-lived. 
The  NLSP can be detected directly at the collider, the decay lifetime provide the information of hidden sector SUSY breaking. If gravitino is not the LSP,  the gravitino can be long-lived 
and may  have impact on big-bang neucleosynthesis. See Fig. \ref{SUSYB}.

\begin{figure}
\includegraphics[width=8cm, angle=0]{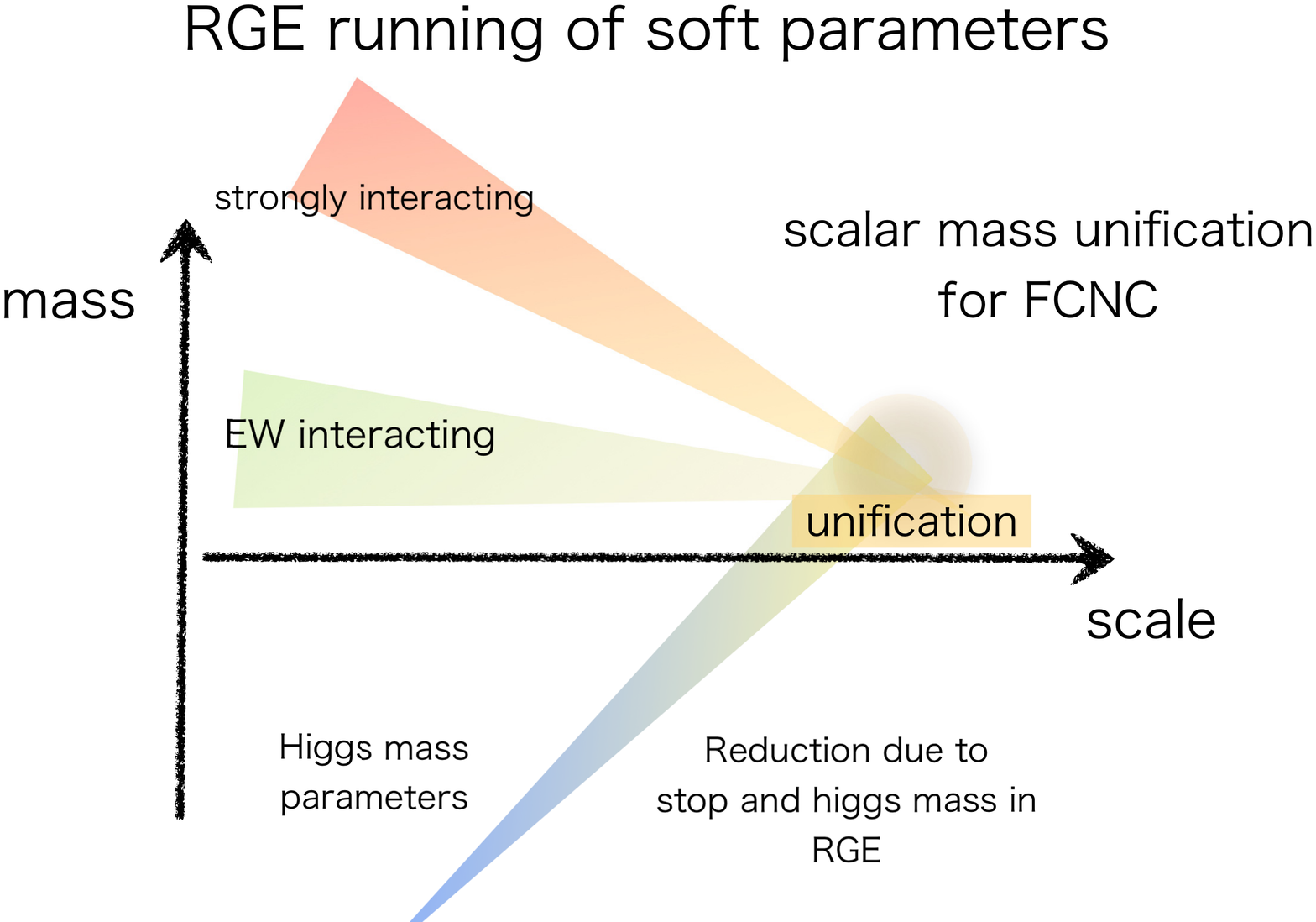}
\includegraphics[width=8cm, angle=0]{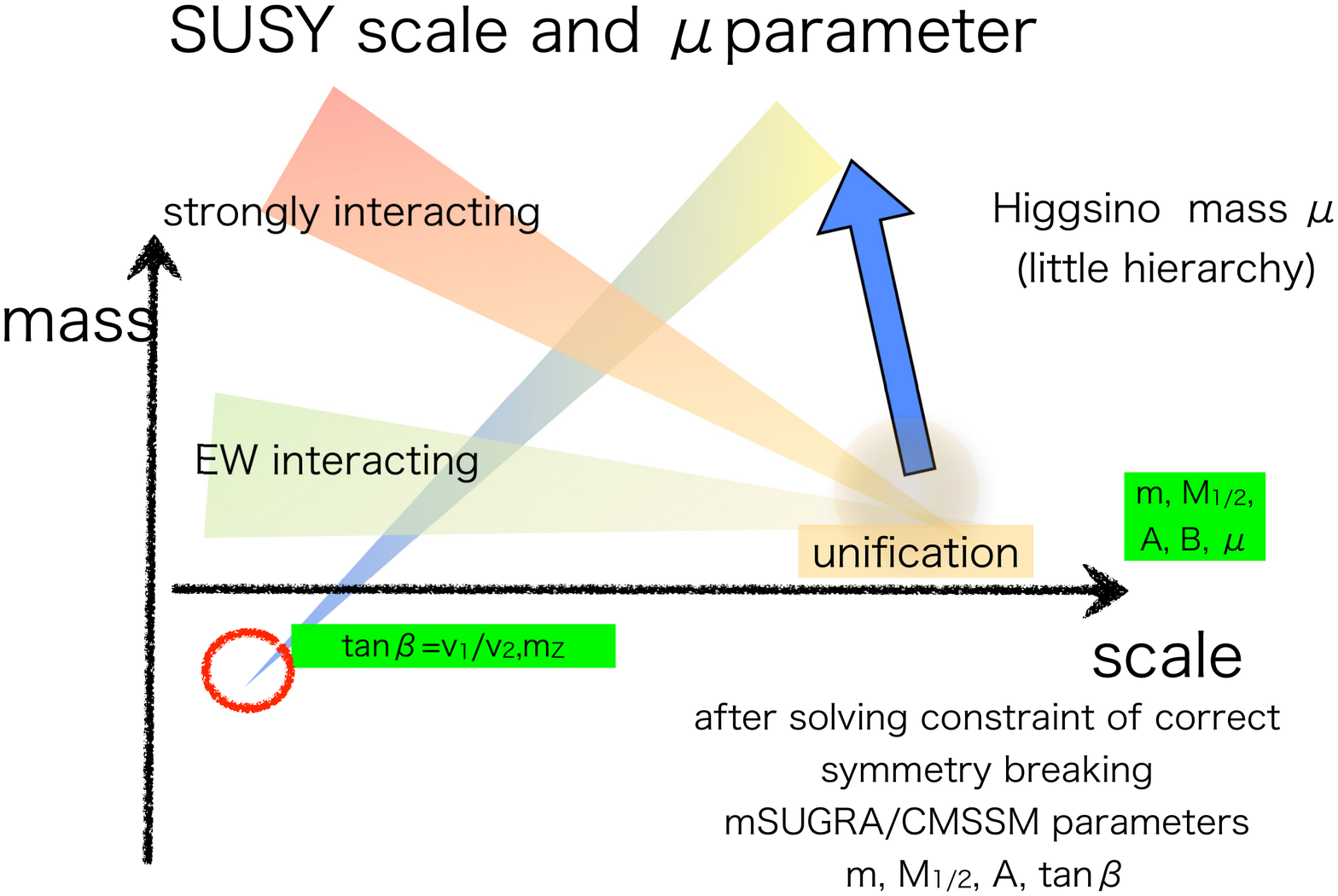}
\caption{Relation between the  MSSM sector and SUSY breaking sector.}\label{radiative}
\end{figure}

The mediation mechanism sets the sparticle mass parameters  at 
the mediation scale, and on-shell masses of the SUSY particles are   obtained by running the RGE equation of the masses down to 
the low energy scale. If the boundary condition is  universal  at $M_{GUT}$,
squark and gluino masses are much heavier than those of electroweakly interacting  superpartners such as sleptons,  
wino, bino  and Higgsinos.  The square of Higgs mass parameter is driven to be 
negative at the weak scale, and Higgsino mass parameter $\mu$  compensates it so that 
the Higgs vev is the correct value. The cancellation between $\mu$  and SUSY breaking parameters at the weak scale is a measure of the  fine turning in the Higgs sector. See Fig. \ref{radiative}. 

\section{Collider search of supersymmetric particles}  
So far,  a proton-proton collider  at CERN, the Large Hadron Collider (LHC),   has 
collected $\sim 30$ fb$^{-1}$ of integrated luminosity 
for each experiment  at 7 to  8~TeV. It will start operation  again  from 2015 aiming for 
300~fb$^{-1}$
at 13~TeV.   

A proton is a composite particle and quarks and gluons in the proton are the 
elementary particles that are involved in the high energy scatting process. 
The momentum of the quarks and gluons are parallel to the beam direction
but the absolute values are not fixed. Therefore the collision system  
is boosted to one of  the beam directions.   The production cross section is 
generally the highest near the threshold. It  reduces gradually with the increase of the  parton collision  energy  $\sqrt{s}$.  The quarks and gluons in the final state are fragmented and 
hadronized into hadrons, forming the jets. Electroweakly interacting particles $W$, $Z$, $\gamma$, leptons and neutrinos  are also produced 
from various  production processes. 
 
\begin{figure}
\centering\includegraphics[width=10cm, angle=0]{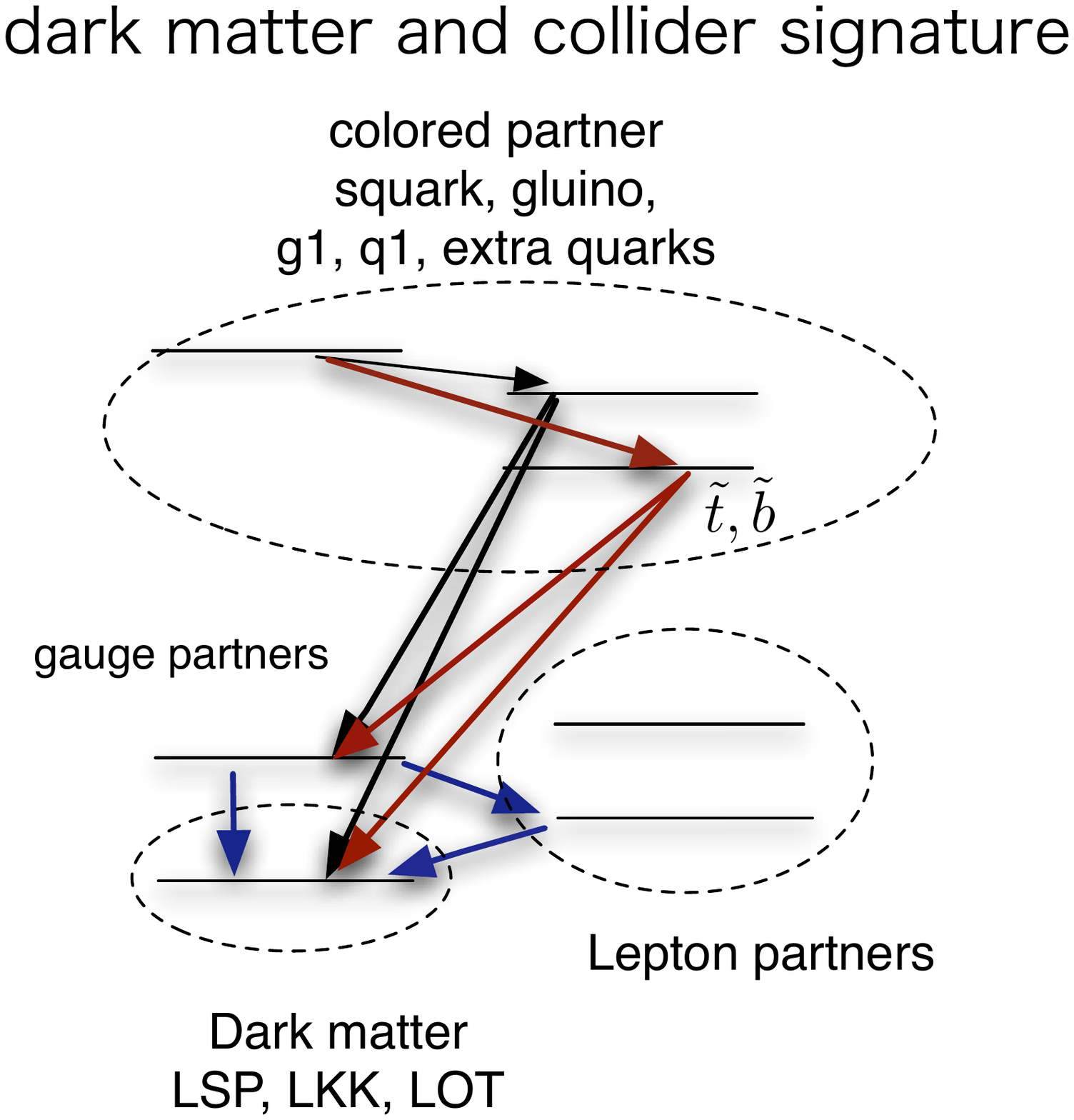}
\caption{The decay pattern of squark and gluino produced at the LHC, and 
particles emitted from the cascade decay chain. The particles in the Little Higgs model with T parity or universal extra dimension model may also give a similar signature.  }\label{sparticle}
\end{figure}

Colored superpaticles are copiously produced at the hadron collider. 
Due to the conserved $R$-parity 
of the MSSM, superpartners are produced in pairs, 
each superpartner  decays to the final state involving another superpartner, 
and at the end of the cascade decay, the LSP appears.  
The LSP is stable. Due to the cosmological constraints, 
it is neutral and color-singlet,  and escapes detection. 
If the mass difference between the superpartners  are large,  the decay product
tends to have high $p_T$. 
In such a case, the LSP, which cannot be detected directly,  is also  relativistic
(See Fig.~ \ref{sparticle}). 
The sum of LSP momentum transverse to the beam direction is balanced
against other visible particles.  Namely, significant  missing transverse momentum
${\bf P}_{Tmiss}$  defined as 
\begin{equation}
{\bf P}_{Tmiss}=-\sum_{i} {\bf p_T}^i_{\rm jet} + \sum_j {\bf p_T}^j_{\rm l}, 
\end{equation}
is a signature of SUSY particle production. 
Another important quantity is the sum of absolute values of the transverse momentum 
\begin{equation}
H_T =\sum_{i}  p^i_{\rm Tjet} + \sum_j  p^j_{\rm Tl}, 
\end{equation}
or the  effective mass 
\begin{equation}
m_{eff} =\sum_{i}  p^i_{Tjet} + \sum_j  p^j_{Tl} + E_{Tmiss}, 
\end{equation}
where $E_{Tmiss }$ is the absolute value of missing transverse momentum.

\begin{figure}
\centering\includegraphics[width=7cm]{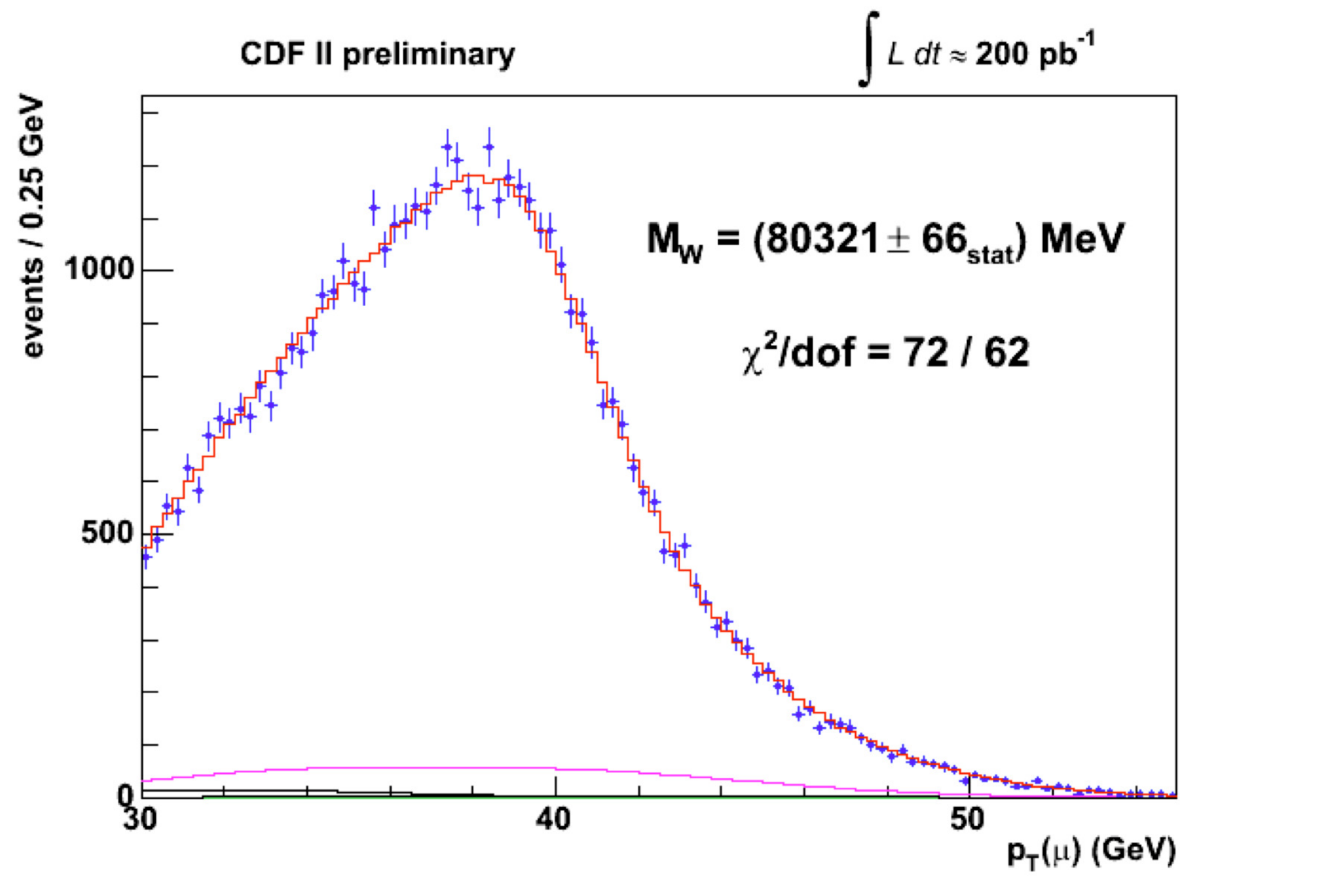}
\caption{Distribution of $p^{\mu}_T$ from the $W$ boson decay measured at CDF experiment at Tevatron.}\label{boson}
\end{figure} 
The $m_{\rm eff}$ distribution peaks at the  sum of the produced particles 
at  the hard process. 
To observe this fact, let us first consider the $p_T$ distribution of  leptons from $W$ boson decay  
produced at  CDF experiment at Tevatron, a $p\bar{p}$ collider at 1.8~TeV. The distribution peaks at 40~GeV, which is a
half of the  $W$ boson mass. See Fig.~\ref{boson}. The feature is easily  understood when we 
calculate the $p_T$  distribution  of spherically decaying $W$ boson boosted to the beam direction, 
\begin{equation}
f(x)dx= \frac{2}{\sqrt{1-x^2}}dx, 
\end{equation}
where $p_T= (m_W/2)\sin\theta= xm_W/2$; 
The distribution strongly peaks at $p_T=m_W/2$ ($\sin\theta=1$) and the structure 
remains even though  $W$ bosons are boosted transversely in the realistic situation, 
because the production cross section is largest near the threshold.   
The fact applies to all production processes at the hadron collider;  the sum of the $p_T$ of the decay products peaks near the parent's mass.  When heavy particles are produced in pairs, 
the sum of the $p_T$  of the decay products  peaks at the sum of the produced 
particle masses. 

\begin{figure}
\centering\includegraphics[width=10cm]{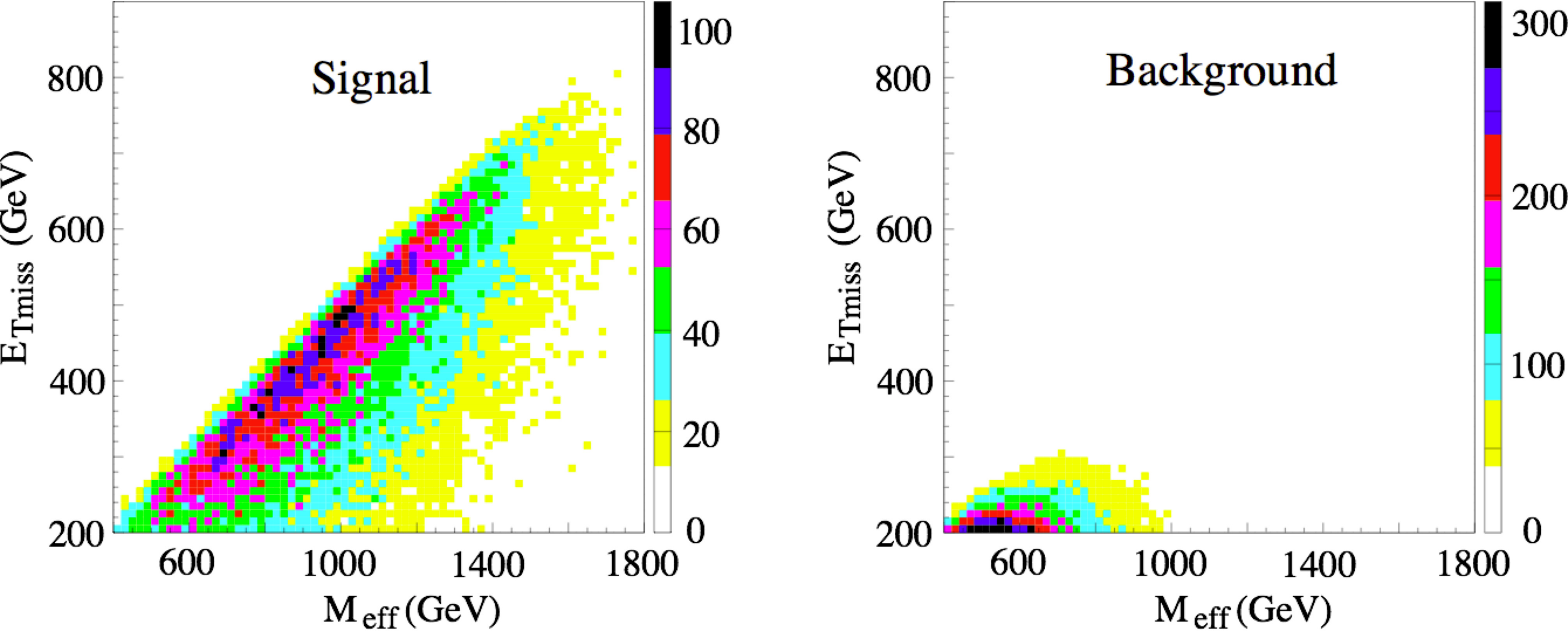}
\caption{Distribution of $m_{eff}$ and $E_{Tmiss}$  of the 
top partner pair production at the LHC followed by the decay into 
top and stable neutral gauge boson (left) compared with the $t\bar{t}$ distribution.  }
\label{sigbag}
\end{figure}
Fig.~\ref{sigbag} compares the distributions of  $T_{-}T_{-}$ and $t\bar{t}$  pair productions.
Here  a hypothetical particles $T_{-}$  
is assumed to decay into $t$ and $B_H$, and  $B_H$ is a neutral stable massive
$U(1)$ gauge boson. The signal contains $t\bar{t}$ and existence of  two $B_H$'s is
observed by the missing transverse momentum of the events, namely, the signal 
is similar to that of superpartner pair production.   
The signal production cross section is  ${\mathcal O}$(1)~pb, while 
the $t\bar{t}$ production cross section is huge at the LHC, 
around 800~pb. 
If the distribution overlaps significantly, the signal is very difficult to be observed. 
However, the signal $m_{eff}$ distribution peaks around  1~TeV and missing momentum 
as close as half of the $M_{eff}$, while the background peaks around $m_{eff}\sim 400$ 
GeV and $E_{Tmiss}\ll M_{eff}/2$. Because of this distribution differences, the $T_{-}$
signature with the production cross section much less than 1~pb may be observed 
at the LHC. 

So far we have been talking about ``inclusive'' quantity. They are defined using 
all objects in an event. We may also select jets or leptons with special features
and use kinematical information to separate signals and 
backgrounds.  Let us consider events with one lepton and some missing momentum. 
The event with one lepton + multiple jets + missing momentum is an important 
signature of superpartner production. However, events involving $W$ boson 
also produce such signatures.  However, the events with $W$ boson can be reduced 
significantly if  we require  that $m_T$ of a lepton and missing $p_T$ is above 100~GeV 
where $m_T$ is defined as 
\begin{equation}
m_T=\sqrt{2 p^l_T E^{Tmiss} }(1-\cos(\Delta\phi(l,p_T))
\end{equation}
The cut significantly reduces the background from the $W$ boson production
to the SUSY process. 

The current bound of the SUSY process is obtained after successful reduction 
of background using the above kinematical variable.  The understanding of background 
distribution 
is quite important, especially the cross section of $W$,$ Z$, $t\bar{t}$ 
with multiple jets must be correctly calculated.  The techniques to 
obtain multiple jets amplitudes with parton shower has been established only 
this century, and current SUSY searches at the LHC  is benefitted by those techniques 
greatly.  The current limit typically excludes squark with mass 1.8~TeV and gluino 
with mass less than 1.4~TeV, if the mass splitting between the LSP and colored SUSY 
particles are large enough. See Fig.~ref{susylimit} for the latest limits. 

\begin{figure}
\centering\includegraphics[width=7cm,angle=-90]{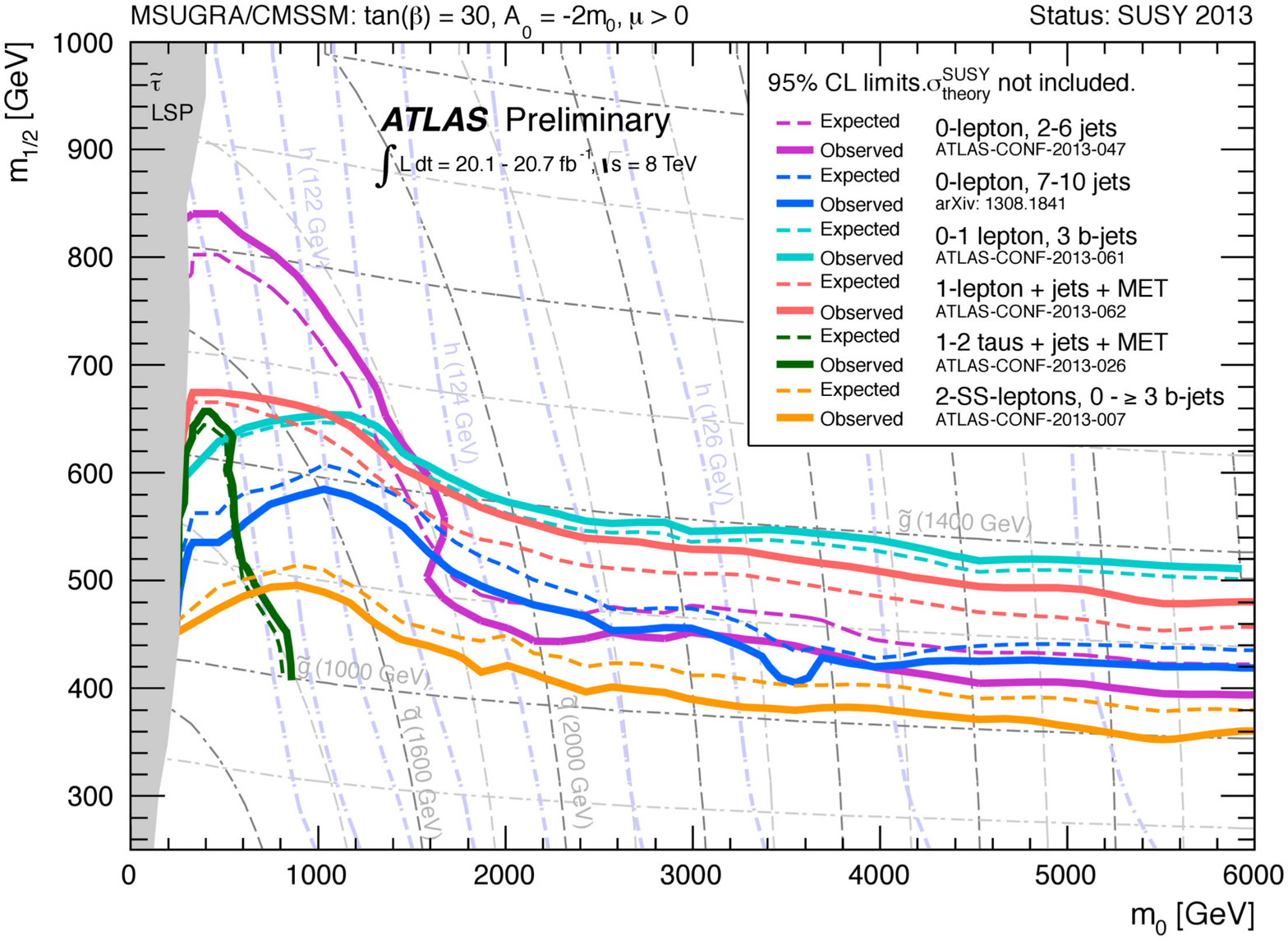}
\caption{Latest mass limit of the MSSM squarks and gluinos shown as a function of GUT 
scale gaugino  mass and scaler mass.  Presented in SUSY2013. }\label{susylimit}
\end{figure}
 
 \section{Dynamical symmetry breaking and BSM }  
Supersymmetry is not a unique solution of the hierarchy problem. 
Another important class of solutions is  dynamical symmetry breaking 
models.  When a global  symmetry is broken spontaneously, a massless 
scalar modes (Nambu-Goldstone boson) appears, even if  the theory does not have an elementary
Higgs boson.  An important example is chiral symmetry breaking in QCD. 
The QCD Lagrangian 
has  $SU(2)_L \times SU(2)_R$ symmetry when quark masses are ignored. The symmetry   is spontaneously broken  
to $SU(2)_V$ dynamically, and  the Goldstone boson of the symmetry breaking are pions $\pi \sim \bar{q}_i \gamma_5 q'$,
and $\langle \bar{q} q \rangle$ has  non-zero vev. 

The pion  has the same charge as the Goldstone boson in the Higgs sector. 
 Therefore, it is natural to consider scale up of the mechanism. The model involves  a set of new quarks $Q$ with  EW charges, but couple to different asymptotic free gauge interactions whose couplings blow up at the scale of EW symmetry breaking.  If $\bar{Q}Q$ condense, 
 the light $\bar{Q}\gamma_5 Q$ states work as the Goldstone bosons of the EW symmetry breaking. 
 This class of the model called Technicolor model. The model has no quadratic divergence because  the massless bound states only appear in the low energy effective theory. 
 
This is an interesting and beautiful  idea, but is not consistent with precision EW observations. 
At LEP, gauge boson two point functions were  precisely 
measured.   Especially the parameter called $S$,  
receives non decoupling contribution from 
$SU(2)$ doublets $Q$ which is colored in the new strong interactions, and 
also necessary charged under $SU(2)\times U(1)$ symmetry in the SM to break 
the gauge symmetry.  Their contribution  appears 
constructively to the gauge two point functions, and therefore the model is tightly constrained. In addition, these models tend to predict a heavy Higgs boson inconsistent with the data. 
  
Another class of models called "composite Higgs models"  allows  a Higgs boson which is  light  but non-elementary.  
In these models,  the Higgs doublet itself is a pseudo Goldstone boson of  some dynamical symmetry breaking.   Though the mechanism of dynamical symmetry breaking is not specified,  the smallness of the mass of the Higgs boson is thought to be  ensured by the  global symmetry of the theory.  The model  requires extension of  the top sector because the top Yukawa coupling 
 violates the desired global symmetry strongly.  The extended top sector is a target of extensive ATLAS and CMS searches.

\section{Extra dimension models }
In the Extra dimension models  the space has more than three dimensions but the 
additional  space dimension is compactified with  a small size $R$ so that we could 
not recognized it easily.  When the extra dimension is flat, the fields in the extra dimension may satisfy the periodic 
boundary condition such as  
\begin{equation}
\phi(x,y)=\phi(x,y+R),
\end{equation}
where $x$ represents four dimensional space time, while $y$ is the 
fifth dimension. 
Under this boundary condition, the wave function is 
expressed as 
\begin{equation}
\psi(x, y) = \psi'(x) \exp (ip_5 y), 
\end{equation}
where $p_5 R = 2\pi n $ ($n$ is an integer). 
This leads to an equation of motion of a free particle propagating in the the fifth dimension, 
\begin{equation}
E_n^2=p^2+p^2_5 = p^2+ (2\pi)^2 \left(\frac{n}{R}\right)^2 .
\end{equation}
Namely, the model predicts an infinite tower of particles of the four dimensional effective 
theory, which corresponds to different values of the discrete momenta in  the fifth direction. 

The coupling of the fifth dimension related with the couplings in the four dimensional effective 
theory in non-trivial manner.   A simple example is the gauge coupling of the fifth dimensional 
theory and the four dimensional  effective theory, 
\begin{equation}
\int d^4x dx_5 \frac{1}{g^2_5} F_{\mu\nu}F^{\mu\nu}
\rightarrow \int d^4x  \frac{1}{g^2_4} F_{\mu\nu}F^{\mu\nu} ,
\end{equation}
where $g_4=g_5/\sqrt{R}$. Larger the size of the fifth dimension is, $g_4$ becomes small. 
This is also true for gravitational interactions. The four dimensional gravitational 
interaction may be small because the size of extra dimension is large. The Large extra dimension 
model tried to solve the fine tuning problem by making true Phanck scale in the higher dimensional theory much smaller than the $M_{pl}$.  

The extra dimension may not be flat. In the RS model, the 
fifth dimension has non-trivial metric as follows:
\begin{equation}
ds^2=e^{-2\sigma(\phi)}\eta_{\mu\nu} dx^{\mu}dx^{\nu}
+r_c^2 d\phi^2 , 
\end{equation}
where $\phi=0$ and $\pi$ is the boundary of the fifth dimension.
The gravity action in the bulk is expressed as 
\begin{equation}
S_{gravity}=\int d^4x \int^{\pi}_{-\pi} d\phi \sqrt{-G}{-\Lambda + 2M^3 R }, 
\end{equation}
when the $\sigma(\phi) $  is expressed as 
\begin{equation}
\sigma(\phi)=r_c\vert\phi\vert \sqrt{\frac{-\Lambda}{24M^3}},
\end{equation}
provided appropriate fine tuning of the boundary actions. 

The  geometry allows us to control the masses of SM particles. 
If the Higgs boson is  at $\phi=\pi$ boundary (which is called 
visible brane), the kinetic term  is expressed as 
\begin{equation}
S_{vis}=\int d^4 x \sqrt{-\bar{g}_{vis} } e^{-4kr_c\pi}
\times \left\{
g^{\mu\nu}_{vis} e^{2kr_c\pi}
D_{\mu} H^{\dagger}D_{\nu} H -\lambda (\vert H \vert ^2 -v_0^2)^2 
\right\}. 
\end{equation}
The mass term receives the suppression factor 
of  $e^{-kr_c\pi}$ after rescaling the Higgs field so that they have canonical kinetic terms.  By adjusting parameters one can easily obtain the mass of the SM 
particle of the oder of the EW scale while  all 
parameters of the fundamental fifth dimensional Lagrangian  are of the order of $M_{pl}$
 without fine turning. 

The model predicts towers of KK particles with mass of the 
order of $\Lambda_{\phi}= \sqrt{6}M_{pl} e^{-kr_c\pi} $ for the particles
living in the fifth dimension(bulk). A popular set up of the model is that 
all the SM fermions are  the zero  mode of the particles 
living in the bulk, and the Higgs boson lives in the IR brane. 
Mass term of the fifth dimensional Lagrangian of the 
SM model matters control the profile of the fields in the bulk. One can adjust the mass so that 
light (heavy) quarks and lepton have small (large)  overlap 
with the IR brane so that Yukawa couplings in the four dimensional effective 
Lagrangian is realized without introducing too much hierarchy among the 
interactions between the Higgs boson and the bulk fermions. 
There are on-going search of the KK gauge bosons and KK fermions 
at the LHC, however, FCNC constraints require
$\Lambda_\phi> 10$~TeV already, and it is unlikely that these new particles 
will be  found at the LHC. 

\section{Suggested reading}
To those who is interested in Supersymmetry, a good review for start with is 
  S.~P.~Martin,``A Supersymmetry primer,''
  In *Kane, G.L. (ed.): Perspectives on supersymmetry II* 1-153
  [hep-ph/9709356]. For A review of composite Higgs model, I suggest 
 R.~Contino, ``The Higgs as a Composite Nambu-Goldstone Boson,''
  arXiv:1005.4269 [hep-ph].
    \end{document}